\documentclass [12pt,eqsecnum,amsfonts,amssymb,aps]{revtex4}

\input epsf

\usepackage[usenames]{color}
\usepackage{graphicx}
\usepackage{bm}
\usepackage{rotating}

\newcommand{\beq}{\begin{equation}}
\newcommand{\eeq}{\end{equation}}
\newcommand{\beqs}{\begin{eqnarray}}
\newcommand{\eeqs}{\end{eqnarray}}

\begin{document}

\baselineskip 6.0mm

\title{Asymptotic Behavior of Spanning Forests and Connected Spanning Subgraphs
  on Two-Dimensional Lattices} 

\author{Shu-Chiuan Chang$^a$ and Robert Shrock$^b$}

\affiliation{(a) \ Department of Physics, National Cheng Kung University,
Tainan 70101, Taiwan} 

\affiliation{(b) \ C. N. Yang Institute for Theoretical Physics and
Department of Physics and Astronomy \\
Stony Brook University, Stony Brook, NY 11794, USA }

\begin{abstract}

  We calculate exponential growth constants $\phi$ and $\sigma$ describing the
  asymptotic behavior of spanning forests and connected spanning subgraphs on
  strip graphs, with arbitrarily great length, of several two-dimensional
  lattices, including square, triangular, honeycomb, and certain heteropolygonal
  Archimedean lattices. By studying the limiting values as the strip widths get
  large, we infer lower and upper bounds on these exponential growth constants
  for the respective infinite lattices. Since our lower and upper bounds are
  quite close to each other, we can infer very accurate approximate values for
  these exponential growth constants, with fractional uncertainties ranging
  from $O(10^{-4})$ to $O(10^{-2})$.  We show that $\phi$ and $\sigma$, are
  monotonically increasing functions of vertex degree for these lattices.

\end{abstract}

\maketitle


\pagestyle{plain}
\pagenumbering{arabic}


\section{Introduction}
\label{intro_section}

Let $G=(V,E)$ be a graph defined by its vertex and edge sets $V$ and $E$.  Let
$n(G)=|V|$, $e(G)=|E|$, and $k(G)$ denote the number of vertices (=sites),
edges (= bonds), and connected components of $G$, respectively.  The degree
$\Delta$ of a vertex $v \in V$ is the number of edges that are incident on $v$.
A graph with the property that all vertices have the same degree is denoted a
$\Delta$-regular graph.  A subgraph of $G$ is defined as a graph with vertex
and edge sets that are subsets of $V$ and $E$. For a given $G$, it is of
interest to enumerate the number of subgraphs of a specific type. A spanning
subgraph is defined as a subgraph that contains all of the vertices of $G$ and
a subset of the edges of $G$.  In the construction of a spanning subgraph there
is a two-fold choice for each edge of $G$, namely whether it is present or
absent, so the number of spanning subgraphs of $G$ is $N_{SSG}(G)=2^{e(G)}$.
We shall restrict here to initial graphs $G$ that are connected and do not have
any loops, i.e., edges that emerge from a given vertex, loop back, and
reconnect to this vertex. Because our focus here is on regular lattice graphs
and their $n\to\infty$ limits, we shall also restrict to graphs that do not have
multiple edges connecting adjacent vertices.  In general, a spanning subgraph
may contain cycles, i.e., paths along edges of the subgraph that are circuits.
Spanning forests on $G$ (abbreviated as SF) are defined as spanning subgraphs
of $G$ that do not contain any cycles.  Note that a spanning forest may consist
of more than one connected component, i.e., are not connected.  We denote the
number of spanning forests of a graph $G$ as $N_{SF}(G)$.  A second set of
subgraphs of $G$ is comprised of connected spanning subgraphs (abbreviated as
CSSG).  We denote the number of these connected spanning subgraphs as
$N_{CSSG}(G)$ and observe that a member of this set may contain cycles.

The numbers of spanning forests and connected spanning subgraphs in a graph $G$
can be calculated as special valuations of the Tutte (also called
Tutte-Whitney) polynomial of $G$, $T(G,x,y)$, defined as
\beq
T(G,x,y) = \sum_{G' \subseteq G} (x-1)^{k(G')-k(G)} \, (y-1)^{c(G')} \ ,
\label{t}
\eeq
where $G'$ is a spanning subgraph of $G$ and $c(G')$ denotes the number of
(linearly independent) cycles on $G'$ \cite{graphtheory,tutte67,bo}. Recall
that we take $k(G)=1$. As is evident directly from the definition (\ref{t}),
the number of spanning forests in $G$ is
\beq
N_{SF}(G) = T(G,2,1)
\label{tg21}
\eeq
and the number of connected spanning subgraphs in $G$ is 
\beq
N_{CSSG}(G) = T(G,1,2) \ . 
\label{tg12}
\eeq
The Tutte polynomial is equivalent to the Whitney rank polynomial
\cite{whitney} and to the Potts model partition function (see 
Appendix \ref{graphtheory_appendix}). 

An interesting problem in graph theory is to calculate the asymptotic behavior
of $N_{SF}(G)$ and $N_{CSSG}(G)$ as $n(G) \to \infty$ for some families of
graphs.  For a wide class of families of graphs, $N_{SF}(G)$ and 
$N_{CSSG}(G)$ grow exponentially rapidly as functions of $n(G)$ for large 
$n(G)$.  This is true, in particular, for lattice graphs. It is thus
natural to define exponential growth constants describing this asymptotic
behavior: 
\beq
\phi( \{ G \} ) = \lim_{n(G) \to \infty} [N_{SF}(G)]^{1/n(G)}
\label{phi}
\eeq
and
\beq
\sigma( \{ G \} ) = \lim_{n(G) \to \infty} [N_{CSSG}(G)]^{1/n(G)} \ , 
\label{sigma}
\eeq
where the symbol $\{ G \}$ denotes the $n(G) \to \infty$ limit of graphs in a
given family. Two simple examples of families of graphs for which one has exact
expressions for $T(G,x,y)$ and hence exact values of $\phi$ and $\sigma$ are
$n$-vertex tree graphs, $T_n$, and circuit graphs, $C_n$, for which
$T(T_n,x,y)=x^{n-1}$ and $T(C_n,x,y)=y+\sum_{j=1}^{n-1}x^j$.  Hence, in the $n
\to \infty$ limits of the $T_n$ and $C_n$ graphs, $\phi=2$ and $\sigma=1$.
However, the problem of calculating $\phi(\Lambda)$ and $\sigma(\Lambda)$ on
infinite lattices $\Lambda$ with dimension 2 or higher is open.

In this paper we present exact calculations of these exponential growth
constants for spanning forests and connected spanning subgraphs on strips, with
fixed width and length going to infinity, of several types of two-dimensional
lattices, including square, triangular, honeycomb, and certain heteropolygonal
Archimedean lattices. By calculating the limiting values of the exponential
growth constants as functions of strip width for infinite-length strips, we
infer lower and upper bounds on these exponential growth constants for the
respective infinite lattices, denoted generically as $\Lambda$. Our lower and
upper bounds are quite close to each other, which enables us to infer very
accurate approximate values for these exponential growth constants, with
fractional uncertainties ranging from $\sim O(10^{-4})$ to $\sim
O(10^{-2})$. We show that $\phi$ and $\sigma$, are monotonically increasing
functions of vertex degree for these lattices.  Our methods of obtaining lower
and upper bounds on $\phi(\Lambda)$ and $\sigma(\Lambda)$ are similar to those
that we have used in our earlier works \cite{ka3}-\cite{aca} in which we
inferred lower and upper bounds on the exponential growth constants for acyclic
orientations, acyclic orientations with a unique source, and totally cyclic
orientations of directed graphs.  Our results make interesting connections
between statistical physics and mathematical graph theory, since the Tutte
polynomial is equivalent to the partition function of a classical spin model,
namely the Potts model.

Previous studies have focused on lower and upper bounds on $\phi$ on the 
square and/or triangular lattices \cite{mw99}-\cite{mani2012}. 
After the early work 
\cite{mw99}, Ref. \cite{cmnn} obtained the lower and upper bounds
(given, respectively, in Theorem 6.15 and Corollary 5.4 of \cite{cmnn})
\beq
3.64497565 \le \phi(sq) \le 3.74100178 \ .
\label{cmnn_phi_bounds}
\eeq
Ref. \cite{jss_sf} improved these bounds to
\beq
3.698573 \le \phi(sq) \le 3.73264
\label{jss_phi_sq_bounds}
\eeq
or equivalently, $1.307947 \le \ln[\phi(sq)] \le 1.317115$ (from Eqs. (7.32a)
and (2.41) in \cite{jss_sf}), where the lower bound is inferred from the 
monotonicity of $\phi$ values for infinite-length, finite-width lattice 
strips. Ref. \cite{garijo2014} obtained the bounds (given in Theorem 5.3 of 
\cite{garijo2014}) 
\beq
3.65166 \le \phi(sq) \le 3.73635 \ .
\label{garijo_phi_bounds}
\eeq
A more stringent upper bound was presented in \cite{mani2012}, namely 
\beq
\phi(sq) \le 3.705603 \ . 
\label{mani_phi_upper_bound}
\eeq
Our results in this paper include a further improvement of the upper bound to 
$\phi(sq) \le 3.699659$, as will be discussed below. 
For the triangular lattice, Ref. \cite{jss_sf} obtained the bounds
\beq 5.479547 \le \phi(tri) \le 5.77546 \ .
\label{jss_phi_tri_bounds}
\eeq
or equivalently, $1.7010224 \le \ln[\phi(tri)] \le 1.75362$ (from Eqs. (7.57a)
and (2.43) in \cite{jss_sf}). Our results in this paper include an improvement
of the upper bound to $\phi(tri) \le 5.494840$.  As noted, we also infer lower
and upper bounds on $\phi(\Lambda)$ for a number of other planar lattices, as
well as lower and upper bounds on $\sigma(\Lambda)$ for these lattices. Our
analysis is thus complementary to earlier work on $\phi(\Lambda)$ in that, by
studying a substantial variety of Archimedean lattices with widely differing
vertex degrees ranging from 3 to 6, we are able to infer the interesting
monotonicity relation given below in Eq. (\ref{monotonic}) for $\phi(\Lambda)$.
The exponential growth constants $\sigma(\Lambda)$ for connected spanning
subgraphs on lattices $\Lambda$ do not seem to have received as much attention
as $\phi(\Lambda)$ for spanning forests. Here again, by obtaining lower and
upper bounds, and resultant approximate values of $\sigma(\Lambda)$ on a
variety of Archimedean lattices, we are able to discern the monotonicity
relation (\ref{monotonic}) for $\sigma(\Lambda)$.  For a given lattice
$\Lambda$, our upper bounds on $\phi(\Lambda)$ and $\sigma(\Lambda)$ approach
respective limiting values more rapidly than our lower bounds, so we infer that
the exact values of these exponential growth constants are closer to our upper
bounds. This was also the case with our earlier calculations of bounds on
exponential growth constants for acyclic orientations, acyclic orientations
with a unique source, and totally cyclic orientations for directed lattices in
Refs. \cite{ac,aca}.

This paper is organized as follows. In Section 
\ref{method_section} we discuss our methods for inferring lower and upper
bounds on the exponential growth constants $\phi(\Lambda)$ and
$\sigma(\Lambda)$ for infinite lattices $\Lambda$ from
calculations on infinite-length strip graphs of varying widths.
In Sections \ref{phi_sigma_bounds_section} and 
 \ref{phi_sigma_values_section} we present our results on these lower and 
upper bounds for $\phi(\Lambda)$ and $\sigma(\Lambda)$ and approximate values
of these exponential growth constants 
for various lattices $\Lambda$. Our conclusions are given in Section
\ref{conclusion_section}.  Some graph theory background is included in Appendix
\ref{graphtheory_appendix}.


\section{Calculational Methods}
\label{method_section} 

In this section we explain a method that we use to infer lower and upper bounds
on the exponential growth constants $\phi(\Lambda)$ and $\sigma(\Lambda)$. 
Our method is the same as the one that we have used in previous work 
\cite{ka3,ac,aca} to infer
lower and upper bounds on exponential growth constants for other
graph-theoretic quantities, such as acyclic orientations, $\alpha(\Lambda)$, 
acyclic orientations with a unique source vertex, $\alpha_0(\Lambda)$, and
totally cyclic orientations, $\beta(\Lambda)$, of directed graphs, and so we
refer the reader to these previous works for further details. 

We consider a family of strip graphs of a given type of lattice $\Lambda$
(square, triangular, honeycomb, etc.) of fixed transverse width $L_y$ and
arbitrarily great length $L_x$ with certain boundary conditions.  As indicated,
the longitudinal and transverse directions on the strip are taken to be in the
$x$ and $y$ directions, respectively. (No confusion should result in the use of
the symbols $x$ and $y$ as arguments of the Tutte polynomial $T(G,x,y)$; the
context will always make the meaning clear.) This is a recursive family, in the
sense of Ref. \cite{bds}. We shall indicate the infinite-length limit of a
width-$L_y$ strip graph of the lattice $\Lambda$ with specified transverse
boundary conditions as $\{ \Lambda, (L_y)_{BC_y} \times \infty \}$.  We make
use of a property of the Tutte polynomials for a strip graph of this type, 
namely that it is a sum of certain coefficients multiplied by powers of
various functions, generically denoted $\lambda$, depending on $\Lambda$,
$L_y$, and the boundary conditions, but not on $L_x$.  The powers to which
these $\lambda$ functions are raised are given by the length, $L_x$, of the
strip. As $L_x \to \infty$, the $\lambda$ function with the
largest magnitude dominates the sum.  Henceforth, we will denote this simply as
$\lambda$ for a given strip. From our calculations of Tutte polynomials 
for strip graphs of various lattices, we know this dominant $\lambda$
function in each case. Some of our calculations are in 
\cite{ka3},\cite{a}-\cite{zttor}; some others are listed in \cite{jemrev}. 
Thus, for the infinite-length limit of a given
finite-width strip of some lattice $\Lambda$, to calculate $\phi$ or $\sigma$,
we only need this dominant $\lambda$ function. This is a significant
simplification, since for a general graph, the calculation of the Tutte
polynomials $T(G,1,2)$ and $T(G,2,1)$ are $\#{\rm P}$ hard \cite{jvw,welsh}. 
This $\lambda$ function, and
hence the results for $\phi$ and $\sigma$, are independent of the boundary
condition (free, periodic, or M\"obius) in the longitudinal direction, but do
depend on the boundary condition in the transverse direction, denoted $BC_y$.
We therefore denote the results as $\phi( \Lambda, (L_y)_{BC_y} \times \infty)$
and $\sigma( \Lambda, (L_y)_{BC_y} \times \infty)$.  These are given by the
following limits:
\beq
\phi(\Lambda,(L_y)_{BC_y} \times \infty) = \lim_{n \to \infty}
[N_{SF}(\Lambda,(L_y)_{BC_y} \times L_x)]^{1/n} =
[\lambda(\Lambda,L_y,BC_y)(2,1)]^{\frac{1}{c_\Lambda L_y}}
\label{phi_infstrip}
\eeq
and
\beq
\sigma(\Lambda,(L_y)_{BC_y} \times \infty) = \lim_{n \to \infty}
[N_{CSSG}(\Lambda,(L_y)_{BC_y} \times L_x)]^{1/n} =
[\lambda(\Lambda,L_y,BC_y)(1,2)]^{\frac{1}{c_\Lambda L_y}} \ , 
\label{sigma_infstrip}
\eeq
where the arguments $(2,1)$ and $(1,2)$ are the arguments of the respective 
Tutte polynomial; $c_\Lambda$ is a constant depending on $\Lambda$, with 
$c_{sq}=c_{tri}=1$, $c_{hc}=2$, etc.; and for brevity of notation, we set 
$n(G) \equiv n$. 

Next, for each type of lattice $\Lambda$, we study the dependence of the
exponential growth constants for the finite-width, infinite-length strips,
$\phi(\Lambda,(L_y)_{BC_y} \times \infty)$ and $\sigma(\Lambda,(L_y)_{BC_y}
\times \infty)$, on the strip width.  For both types of transverse boundary
conditions (free and periodic) and for all lattices $\Lambda$ considered here,
we show that these are monotonically increasing functions of the strip
width. This provides strong support for the inference that these are lower
bounds on the respective exponential growth constants $\phi(\Lambda)$ and
$\sigma(\Lambda)$ the infinite lattices $\Lambda$.  Furthermore, a
consequence of this monotonicity property is that, for a given transverse
boundary condition $BC_y$, the values of $\phi(\Lambda,(L_y)_{BC_y} \times
\infty)$ and $\sigma(\Lambda,(L_y)_{BC_y} \times \infty)$ on the strip with the
greatest width are the best lower bounds, for this set of transverse boundary 
conditions, on the respective values of 
$\phi(\Lambda)$ and $\sigma(\Lambda)$. Since for 
the periodic (P) transverse boundary conditions the strips have no transverse
boundary, one expects that these would yield the best lower bounds, and our
results are in accord with this expectation. Thus, for the infinite lattice
$\Lambda$, we infer
\beq
\xi(\Lambda) > \xi(\Lambda,[(L_y)_{max}]_{PBC_y} \times \infty) \quad {\rm for}
\ \xi= \phi, \ \sigma
\label{ximin}
\eeq

A measure of the rapidity with which the values of $\phi$ and $\sigma$ on
finite-width, infinite-length strips of a lattice $\Lambda$ approach their
infinite-width limits is provided by the ratio of values of each respective
exponential growth constant on strips of width $L_y$ and $L_y-1$.  We denote
this ratio as 
\beq
R_{\xi,\Lambda,(L_y+1)/L_y,BC_y} \equiv 
\frac{\xi(\Lambda,(L_y+1)_{BC_y} \times \infty)}
     {\xi(\Lambda,(L_y)_{BC_y} \times \infty)} \quad 
{\rm for} \ \xi=\phi, \ \sigma \ . 
\label{ralpha}
\eeq
As will be evident from our results for various lattices, 
even for modest values of the strip widths, these ratios approach
very close to unity, showing the rapid approach to the $L_y \to \infty$ limit.

We proceed to discuss upper bounds on these exponential growth constants $\phi$
and $\sigma$.  Our analysis here is similar to our earlier analyses in
\cite{ka3}-\cite{aca} for the other set of exponential growth constants
$\alpha$, $\alpha_0$, and $\beta$.  We first prove a useful inequality. For
this purpose, we begin by considering lattice strip graphs with width $L_y=2^p$
for some (positive) integer power $p$.  This inequality applies to a generic
exponential growth constant denoted $\xi$ for the Tutte polynomial of a family
of lattice strip graphs for $x \ge 0$ and $y \ge 0$, where $\xi$ is defined as
\beq
\xi( \{ G \},x,y) = \lim_{n(G) \to \infty} [T(G,x,y)]^{1/n(G)} \ .
\label{taudef}
\eeq
For our applications here, $\xi=\phi$ with $(x,y)=(2,1)$ or $\xi=\sigma$ with
$(x,y)=(1,2)$.  If an edge $e \in E$ is not a loop or a bridge (see
Appendix \ref{graphtheory_appendix} for definitions), then the Tutte polynomial
satisfies the deletion-contraction relation
\beq
T(G,x,y) = T(G-e,x,y) + T(G/e,x,y) \ ,
\label{dcr}
\eeq
where $G-e$ denotes $G$ with the edge $e$ deleted and $G/e$ is the result of
deleting the edge $e$ from $G$ and identifying the vertices that had been
connected by this edge. Applying this deletion-contraction relation repeatedly
yields a set of inequalities for the dominant $\lambda$ function for the two
cases of interest here, $(x,y)=(2,1)$ and $(x,y)=(1,2)$. If one compares the
Tutte polynomial for an $4 \times L_x$ strip graph with the Tutte polynomial
for a (disconnected) graph consisting of two copies of an $2 \times L_x$ strip,
then the former has $L_x$ more edges, whose deletion produces the the latter
two graphs. By iterative application of the deletion-contraction theorem, one
can then relate the free strip of width $L_y$ to the graph consisting of two
free strips each of width $L_y/2$. Henceforth, for definiteness, we specialize
to strips of the square lattice (and use the symbol ``F'' for free transverse
boundary conditions).  With appropriate changes, our results apply to strips of
other lattices also.  We then have the series of inequalities, for
$(x,y)=(2,1)$ or $(x,y)=(1,2)$:
\beqs
\lambda_{sq,1,F}(x,y) &&\le [\lambda_{sq,2,F}(x,y)]^{1/2} \le
[\lambda_{sq,4,F}(x,y)]^{1/4} \le [\lambda_{sq,8,F}(x,y)]^{1/8}
\cr\cr
&& \le ... \le \lim_{L_y \to \infty} [\lambda_{sq,L_y,F}(x,y)]^{1/L_y} \ .
\label{ineqa}
\eeqs
Let us focus on one of these inequalities, namely 
$[\lambda_{sq,2,F}(x,y)]^{1/2} \le [\lambda_{sq,4,F}(x,y)]^{1/4}$
The others can be treated in a similar manner. Here 
and below, it is understood that $(x,y)=(2,1)$ for $\phi$ or 
$(x,y)=(1,2)$ for $\sigma$. Since $[\lambda_{sq,2,F}(x,y)]^{L_x}$ is the
dominant $\lambda$ function for the $2 \times L_x$
strip, it determines the corresponding $\phi$ or $\sigma$ in the 
limit of infinite strip length, while $[\lambda_{sq,4,F}(x,y)]^{L_x}$ 
similarly gives the $\phi$ or $\sigma$ function for
infinite-length limit of the the $4 \times L_x$ strip.
Now compare two $L_y=2$ strips with a $L_y=4$ strip. The former strips have 
$L_x$ fewer
edges than the latter, so the Tutte polynomial of the former is smaller than
that of the latter, since the coefficients of the Tutte polynomial (in terms of
variables $x$ and $y$) are positive. That is,
$[\lambda_{sq,2,F}(x,y)]^{2L_x} \le [\lambda_{sq,4,F}(x,y)]^{L_x}$.
This completes the proof of the inequality. By the same type of argument, 
it follows, for example, that
\beqs
\lambda_{sq,1,F}(x,y) &&\le [\lambda_{sq,3,F}(x,y)]^{1/3} \le
[\lambda_{sq,6,F}(x,y)]^{1/6} \le [\lambda_{sq,12,F}(x,y)]^{1/12}
\cr\cr
&&\le ... \le \lim_{L_y \to \infty} [\lambda_{sq,L_y,F}(x,y)]^{1/L_y} \ ,
\label{ineqb}
\eeqs
where here $L_y=3 \cdot 2^s$, where $s$ is a non-negative integer. Other
corresponding inequalities with larger values of $L_y$ follow in the same way.
By a similar argument, one can prove that, with $(x,y)=(2,1)$ or $(x,y)=(1,2)$, 
\beq
\lambda_{sq,L_y,F}(x,y) \le \lambda_{sq,L_y,P}(x,y) \ .
\label{ineqc}
\eeq

Now recall the sequence of inequalities (\ref{ineqa}). The limit as 
$L_y \to \infty$ yields the value of the exponential growth constant on the 
infinite square lattice. Now 
\beq
[\lambda_{sq,L_y,F}(x,y)]^{1/L_y} < [\lambda_{sq,L_y+1,F}(x,y)]^{1/(L_y+1)} \ , 
\label{ineqd}
\eeq
or equivalently, 
\beq
\lambda_{sq,L_y,F}(x,y) < [\lambda_{sq,L_y+1,F}(x,y)]^{L_y/(L_y+1)} \ . 
\label{ineqe}
\eeq
Thus, 
\beq
[\lambda_{sq,L_y+1,F}(x,y)]^{1/(L_y+1)} < 
\frac{\lambda_{sq,L_y+1,F}(x,y)}
      {\lambda_{sq,L_y,  F}(x,y)} \ .
\label{ineqf}
\eeq
From our explicit calculation, we find that
\beq
\frac{\lambda_{sq,8,F}(x,y)}{\lambda_{sq,7,F}(x,y)} <
\frac{\lambda_{sq,7,F}(x,y)}{\lambda_{sq,6,F}(x,y)}
< ... < \frac{\lambda_{sq,3,F}(x,y)}{\lambda_{sq,2,F}(x,y)}
      < \frac{\lambda_{sq,2,F}(x,y)}{\lambda_{sq,1,F}(x,y)} \ .
\label{ineqg}
\eeq
This leads us to infer that the ratio
$\lambda_{sq,L_y+1,F}(x,y)/\lambda_{sq,L_y,F}(x,y)$ serves as an
upper bound for $\phi(sq)$ if $(x,y)=(2,1)$ and for $\sigma(sq)$ if
$(x,y)=(1,2)$. We thus infer the inequalities
\beq
\phi(\Lambda) < \frac{\lambda_{\Lambda,L_y+1,F}(2,1)}
                     {\lambda_{\Lambda,L_y,F}(2,1)}  \quad {\rm for \
  the \ maximal \ calculated \ value \ of} \ L_y
\label{phi_upperbound}
\eeq
and
\beq
\sigma(\Lambda) < \frac{\lambda_{\Lambda,L_y+1,F}(1,2)}
                     {\lambda_{\Lambda,L_y,F}(1,2)}  \quad {\rm for \
  the \ maximal \ calculated \ value \ of} \ L_y \ . 
\label{sigma_upperbound}
\eeq
A useful measure of the approach to the $L_y \to \infty$ limit is provided by
the ratio of upper bounds for adjacent values of $L_y$, namely 
\beq
R_{sq,\frac{L_y^2}{(L_y-1)(L_y+1)},F}(x,y) \equiv
\frac{[\lambda_{sq,L_y,F}(x,y)]^2}
{\lambda_{sq,L_y-1,F}(x,y) \lambda_{sq,L_y+1,F}(x,y)} \ .
\label{ratio_upper}
\eeq
This is the ratio of adjacent upper bounds. As our explicit calculations show,
this ratio rapidly approaches unity (from above) as the strip width $L_y$
increases. 

Applying the analogous arguments for other lattices, we infer the inequalities
corresponding to (\ref{phi_upperbound}) and (\ref{sigma_upperbound}) for these
other lattices.  The ratio of adjacent upper bounds analogous to 
(\ref{ratio_upper}) provides a quantitative
measure of the rapidity of approach to a limit for these other lattices, as for
the square lattice. 


\section{Numerical Values of Lower and Upper Bounds for $\phi(\Lambda)$ 
and $\sigma(\Lambda)$ }
\label{phi_sigma_bounds_section}

In this section we present our results for numerical values of lower and upper
bounds for $\phi(\Lambda)$ on various two-dimensional lattices $\Lambda$.  For
a given lattice $\Lambda$, we denote our lower ($\ell$) and upper ($u$) bounds
as $\xi_\ell(\Lambda)$ and $\xi_u(\Lambda)$, where $\xi=\phi$ or $\sigma$.
With our method, we obtain the lower bounds from strips with periodic
transverse boundary conditions and the upper bounds from strips with free
boundary conditions.

We recall that an Archimedean lattice is defined as a uniform tiling of
the plane with one or more types of regular polygons, such that all vertices
are equivalent, and hence is $\Delta$-regular.  In general, an Archimedean
lattice $\Lambda$ is identified by the ordered sequence of regular polygons
traversed in a circuit around any vertex \cite{gsbook,wn}:
\beq
\Lambda = (\prod p_i^{a_i}) \ , 
\label{arch}
\eeq
where the $i$'th polygon has $p_i$ sides and appears $a_i$ times contiguously
in the sequence (it can also occur non-contiguously). As in \cite{wn}, we
denote the sum of the numbers $a_i$ in the product (\ref{arch}) as $a_{i,s}$.
Of the eleven Archimedean lattices, three are homopolygonal (i.e. each is
comprised of only type of regular polygon), namely the square ($sq$),
triangular ($tri$), and honeycomb ($hc$) lattices. For a homopolygonal
Archimedean lattice composed of $p$-gons, the right-hand side of
Eq. (\ref{arch}) has the simple form $(p^\Delta)$, where, as above, $\Delta$ is
the vertex degree (i.e., lattice coordination number). Thus, in this notation,
the square, triangular, and honeycomb lattices are denoted $(4^4)$, $(3^6$),
and $(6^3)$. The other Archimedean lattices are comprised of more than one type
of regular polygon and hence are termed heteropolygonal.  The heteropolygonal
Archimedean lattices that we will consider here are $(4 \cdot 8^2)$, $(3 \cdot
6 \cdot 3 \cdot 6)$, also known as kagom\'e ($kag$), $(3^3 \cdot 4^2)$, and
$(3^2 \cdot 4 \cdot 3 \cdot 4)$.

In Tables \ref{lowerbounds_phi_sq_table}-\ref{upperbounds_phi_33434_table} and
\ref{lowerbounds_sigma_sq_table}-\ref{upperbounds_sigma_33434_table} we present
our results on lower and upper bounds on $\phi(\Lambda)$ and $\sigma(\Lambda)$
and relevant ratios for the finite-width, infinite-length strips of the various
lattices $\Lambda$. We include results for both free and periodic transverse
boundary conditions on these finite-width, infinite-length strips.  In Table
\ref{phi_sigma_bounds_table} we summarize the best lower and upper bounds that
we have obtained for these lattices. To our knowledge, these are the best
current lower and upper bounds on $\phi(\Lambda)$ for the hc, $(4 \cdot 8^2)$,
$(3 \cdot 6 \cdot 3 \cdot 6)$ (kag), $(3^3 \cdot 4^2)$, and $(3^2 \cdot 4 \cdot
3 \cdot 4)$ lattices and the best upper bounds on $\phi(sq)$ and $\phi(tri)$.
As noted, we are not aware of previous published bounds on $\sigma(\Lambda)$
for these Archimedean lattices. 

There are several important features of our bounds.  First, for each type of
lattice, the lists of ratios of adjacent lower bounds and of adjacent upper
bounds, as functions of strip width $L_y$ show that the lower bounds and the
upper bounds rapidly approach a limiting value.
Second, the upper and lower bounds are very close to each
other. The average of the upper and lower bounds for a given exponential 
growth constant $\xi(\Lambda)$ is 
\beq
\xi_{ave}(\Lambda) = \frac{\xi_u(\Lambda) + \xi_\ell(\Lambda)}{2} \quad  
{\rm for} \ \xi = \phi, \ \sigma \ .
\label{xi_ave}
\eeq
The difference between the average and the upper or lower bound is 
\beq
\delta_{\xi(\Lambda)} = \xi_u(\Lambda) -\xi_{ave}(\Lambda) = 
\xi_{ave}(\Lambda)-\xi_\ell(\Lambda) \ , 
\label{delta_xi} 
\eeq
so the fractional difference is 
\beq
\frac{\xi_u(\Lambda) - \xi_\ell(\Lambda)}{\xi_{ave}(\Lambda)} = 
\frac{2\delta_{\xi(\Lambda)}}{\xi_{ave}(\Lambda)} \quad 
{\rm for} \ \xi=\phi, \ \sigma \ . 
\label{fracdif}
\eeq
These fractional differences (\ref{fracdif}) are very small,
typically varying from $O(10^{-4})$ to $O(10^{-2})$. This is in excellent
agreement with the observed rapid approach of each of these bounds to a
limiting value and consistent with the inference that, in the $L_y \to \infty$
limit, this is a common value, describing the exponential growth constant on
the infinite two-dimensional lattice.

Third, for a given lattice $\Lambda$ and exponential growth constant
$\phi(\Lambda)$ or $\sigma(\Lambda)$, the upper bounds approach a limit more
rapidly than the lower bounds, leading one to infer that the actual value on
the infinite lattice is closer to the upper bound than to the lower bound.  For
example, for the $(4 \cdot 8^2)$ lattice, the ratio of adjacent upper bounds
$R_{tri,\frac{L_y^2}{(L_y-1)(L_y+1)},F}(2,1)$ for the greatest widths in Table
\ref{upperbounds_phi_488_table} is extremely close to 1, being only $1.6 \times
10^{-7}$ greater than 1, while the corresponding ratio of lower bounds is
approximately $1 \times 10^{-3}$ above 1.  Fourth, our numerical results are in
agreement with the three exact duality relations $\phi(sq)=\sigma(sq)$ in
Eq. (\ref{phi_sigma_sq}), $\phi(hc) = [\sigma(tri)]^{1/2}$ in
Eq. (\ref{sigmatri_phihc}), and $\sigma(hc) = [\phi(tri)]^{1/2}$ in
Eq. (\ref{sigmahc_phitri}) for the infinite lattices. Accordingly, we have made
use of these duality relations in Table
\ref{phi_sigma_bounds_table}. Specifically, we have used our upper bound on
$\phi(sq)$, namely $\phi_u(sq) = 3.699659$, as an improvement on the upper
bound $\sigma_u(sq)=3.751149$ obtained directly from the infinite-length,
finite-width strips. Further, we have used the duality relation $\sigma(hc) =
[\phi(tri)]^{1/2}$ together with our upper bound on $\phi(tri)$, namely
$\phi_u(tri)=5.494840085$, to compute an upper bound $\sigma_u(hc) =
2.3441075$, which is more stringent than the upper bound $\sigma_u(hc) =
2.3601982$ obtained directly from infinite-length, finite-width strips of the
honeycomb lattice. Similarly, from duality, we obtain a lower limit
$\phi_\ell(tri)=5.39333314$, which is more stringent than the lower bound
$\phi_\ell(tri)=5.3848542$ obtained directly from the analysis of strips of the
triangular lattice. Moreover, $[\phi_\ell(hc)]^2=7.861223392$ and
$[\phi_u(hc)]^2=7.866798814$, which are better lower and upper bounds than the
respective values $\sigma_\ell(tri)=7.859929$ and $\sigma_u(tri)=7.933005$
obtained directly from the analysis of infinite-length, finite-width strips of
the triangular lattice.  We thus use these improved limits in Table
\ref{phi_sigma_bounds_table}.  An important fourth feature of our results will 
be presented as the relation (\ref{monotonic}) in the next section. 


\section{Approximate Values of $\phi(\Lambda)$ and $\sigma(\Lambda)$ }
\label{phi_sigma_values_section}

Since the fractional differences (\ref{fracdif}) are so small, we can
infer very accurate approximate values $\xi_{app}(\Lambda)$ for these 
exponential growth constants
on the given lattices. A simple way to do this is use the 
averages, $\xi_{ave}(\Lambda)$ together with the differences 
$\delta_{\xi(\Lambda)}$ as a measure of the uncertainty:
\beq
\xi_{app}(\Lambda) = \xi_{ave}(\Lambda) \pm \delta_{\xi(\Lambda)} 
\quad {\rm for} \ \xi = \phi, \ \sigma \ .
\label{xi_approx}
\eeq
This is the procedure that we used in 
Refs. \cite{ac,aca} for certain exponential growth constants describing acyclic
and cyclic orientations of edge arrows on directed lattice graphs.  
We list these approximate values in Table \ref{phi_sigma_approx_table}. 
More complicated analytical methods could also be applied, but this simple
procedure is sufficient as a basis for one of our most important results,
namely that we find that, for all of that lattices considered here, 
\beq
\phi(\Lambda) \ {\rm and} \ \sigma(\Lambda) \ {\rm are \ monotonically \ 
increasing \ functions \ of} \ \Delta,
\label{monotonic}
\eeq
where $\Delta$ is the vertex degree (i.e., coordination number in physics
terminology) of the lattice $\Lambda$.  If we write this $\Delta$ dependence as
an empirical power law, then we find, roughly, that $\phi(\Lambda) \sim
3.7(\Delta/4)$ while $\sigma(\Lambda) \sim 3.7(\Delta/4)^{1.8}$.  By fitting
our upper and lower bounds on the exponential growth constants for
infinite-length, finite-width strips to some assumed functional forms for the
approach to the infinite-width limit (as in \cite{jss_sf} for $\phi(sq)$ and
$\phi(tri)$), we could infer corresponding estimates for the values for the
exponential growth constants, but this is not necessary for our monotonicity
result (\ref{monotonic}).  

The dependences of $\phi(\Lambda)$ and $\sigma(\Lambda)$ on $\Delta$ that we
have found may be compared and contrasted with the $\Delta$-dependence of the
exponential growth constant for the total number of spanning subgraphs of a
lattice $\Lambda$, $N_{SSG}(G)=2^{e(G)}$.  Since $e(G)=n \Delta/2$ for a
$\Delta$-regular lattice graph $G$, it follows that for such graphs
\beq
\lim_{n(G) \to \infty} [N_{SSG}(G)]^{1/n(G)} = 2^{\Delta/2} \ . 
\label{egc_ssg}
\eeq
This is again a monotonically increasing function of $\Delta$, and the property
that the right-hand side of Eq. (\ref{egc_ssg}) increases more rapidly than a
power law as a function of $\Delta$ is consistent with the fact that the
numbers of spanning forests and connected spanning subgraphs are subsets of the
total number of spanning subgraphs.

Another interesting property that we find is that for the homopolygonal
lattices $\Lambda = (p^\Delta)$, the relation $p > \Delta \ \Longleftrightarrow
\ \phi(\Lambda) > \sigma(\Lambda)$ holds. We recall that the case $p=\Delta=4$
is realized for the square lattice, the self duality of which implies that
$\phi(sq)=\sigma(sq)$. Given the connection between $\Delta$ and $p$ for the
homopolygonal Archimedean lattices, this relation is implied by our
monotonicity result (\ref{monotonic}), but it is of interest in its own right.

The analogous relations also hold for exponential growth
constants that we calculated in Refs. \cite{ac,aca}. Recall that the number of
acyclic orientations and totally cyclic orientations of a directed graph $G$
are given by $T(G,2,0)$ and $T(G,0,2)$, respectively, with the corresponding
exponential growth constants 
\beq
\alpha(\{ G \}) = \lim_{n(G) \to \infty} [T(G,2,0)]^{1/n(G)} 
\label{alpha}
\eeq
and
\beq
\beta(\{G\}) = \lim_{n(G) \to \infty} [T(G,0,2)]^{1/n(G)} \ . 
\label{beta}
\eeq
As we noted in \cite{ac,aca}, we found that for the Archimedean lattices that
we considered there, 
\beq
\alpha(\Lambda) \ {\rm and} \ \beta(\Lambda) \ {\rm are \ monotonically \ 
increasing \ functions \ of} \ \Delta \ . 
\label{monotonic2}
\eeq
Furthermore, the relation $p > \Delta \ \Longleftrightarrow \ \alpha(\Lambda) >
\beta(\Lambda)$ holds for the homopolygonal Archimedean lattices, and the
self-duality of the square lattice yields the relation $\alpha(sq)=\beta(sq)$.


\section{Conclusions}
\label{conclusion_section}

In this paper we have calculated the exponential growth constants $\phi$ and
$\sigma$ describing the asymptotic growth of the numbers of spanning forests
and of connected spanning subgraphs, respectively, for finite-width,
infinite-length strips of several different two-dimensional lattices $\Lambda$.
From our calculations, we have inferred lower and upper bounds on these
exponential growth constants $\phi(\Lambda)$ and $\sigma(\Lambda)$ for the
respective infinite lattices $\Lambda$.  Our bounds from calculations on
infinite-length, finite-width lattice strips converge rapidly even for modest
values of strip widths.  Since our lower and upper bounds are quite close
to each other, we can infer obtain quite accurate approximate values for these
exponential growth constants.  Our results show that 
$\phi(\Lambda)$ and $\sigma(\Lambda)$ are monotonically increasing functions of
vertex degree for these lattices. An interesting aspect of our work is the 
connection that is makes between statistical mechanics and mathematical graph 
theory, reflecting the fact the Tutte polynomial is equivalent to the 
partition function of a classical spin model, namely the Potts model.


\begin{acknowledgments}

This research was supported in part by the Taiwan Ministry of Science and
Technology grant MOST 103-2918-I-006-016 (S.-C.C.) and by
the U.S. National Science Foundation grant No. NSF-PHY-1620628 and 
NSF-PHY-1915093 (R.S.).

\end{acknowledgments}


\begin{appendix}


\section{Some Graph Theory Background}
\label{graphtheory_appendix}

In this appendix we include some graph theory background relevant for our
analysis in the paper (for further details, see, e.g., \cite{graphtheory}).  As
in the text, let $G=(V,E)$ be a graph defined by its vertex and edge sets $V$
and $E$.  Let $n(G)=|V|$, $e(G)=|E|$, and $k(G)$ denote the number of vertices
(=sites), edges (= bonds), and connected components of $G$. We restrict to
connected $G$. A loop is defined as an edge that connects a vertex to itself,
and a bridge (co-loop) is defined as an edge that has the property that if it
is deleted, then this increases the number of components in the resultant
graph, relative to the number of components in the initial graph that contained
the bridge. As noted in the text, since our primary application is to regular
lattices, we restrict to graphs $G$ without loops. A spanning subgraph of $G$,
denoted $G'$, is a graph with the same vertex set $V$ and a subset of the edge
set $E$, i.e., $G'=G'(V,E')$ with $E' \subseteq E$. The Tutte polynomial
$T(G,x,y)$ \cite{tutte67} is defined in Eq. (\ref{t}) in the text.  
The numbers of spanning forests and connected spanning subgraphs in $G$,
denoted $N_{SF}(G)$ and $N_{CSSB}(G)$ respectively, are valuations of
$T(G,x,y)$ given by Eqs. (\ref{tg21}) and (\ref{tg12}). The corresponding
exponential growth constants describing the asymptotic behavior of $N_{SF}(G)$
and $N_{CSSG}(G)$ are given in Eqs. (\ref{phi}) and (\ref{sigma}).
From the definition (\ref{t}), it is clear that $T(G,x,y)$ is a polynomial in
the two variables $x$ and $y$, so one can write it as 
\beq
T(G,x,y) = \sum_{i,j} t_{ij} \, x^i \, y^j \ , 
\label{tij}
\eeq
where the $t_{ij}$ can be determined from (\ref{t}). A basic property of
$T(G,x,y)$ that we use in the text is that the nonzero $t_{ij}$ are positive
(integers) \cite{graphtheory,tutte67}.  

Let $G_{pl}$ be a planar graph. Recall that the planar dual, $G_{pl}^*$, of
$G_{pl}$ is defined by a 1-1 correspondence between the vertices (resp. faces) 
of $G_{pl}$ and the faces (resp. vertices) of $G_{pl}^*$.  The Tutte polynomial
satisfies the duality relation 
\beq
T(G_{pl},x,y) = T(G_{pl}^*,y,x) \ . 
\label{tdual}
\eeq
It follows from this duality relation (\ref{tdual}) and the relations 
(\ref{tg21}) and (\ref{tg12}) that
\beq
N_{SF}(G_{pl}) = N_{CSSG}(G_{pl}^*) \ . 
\label{tg21_dual_tg12}
\eeq

Let us denote the number of faces of a planar graph $G_{pl}$ as $f(G_{pl})$ and
recall the Euler relation for a planar graph $G_{pl}$,
\beq
f(G_{pl})-e(G_{pl})+n(G_{pl})=2 \ . 
\label{euler}
\eeq
From the duality relation, it follows that $n(G_{pl}^*)=f(G_{pl})$. 
For $\Delta$-regular graphs $G$, 
\beq
e(G) = \frac{\Delta(G) \, n(G)}{2} \ .
\label{egdelta}
\eeq
For a $\Delta$-regular planar graph $G_{pl}$ we define the ratio
\beq
\nu_{ \{ G_{pl} \} } \equiv \lim_{n(G_{pl}) \to \infty} 
\frac{n(G_{pl}^*)}{n(G_{pl})} = \frac{\Delta(G_{pl})}{2}-1 \ , 
\label{nu_g}
\eeq
where we have used Eq. (\ref{egdelta}) in the last equality in (\ref{nu_g}). 
Note that 
\beq
\nu( \{ G_{pl} \}) = \frac{1}{ \nu( \{ G_{pl}^* \} ) } \ . 
\label{nunuinverse}
\eeq
Specifically, $\nu(sq) = 1$ and $\nu(tri) = 1/\nu(hc) = 2$.  The results
$\nu(sq) = 1$ and that $\nu(tri) = 1/\nu(hc)$ follow from property that the
square lattice is self-dual and the triangular and honeycomb lattices are
planar duals of each other.  From Eq. (\ref{tg21_dual_tg12}), it follows that
if a planar graph is self-dual, indicated as $G_{pl.,sd.}$, then
\beq
N_{SF}(G_{pl.,sd.})=N_{CSSG}(G_{pl.,sd.}) \ , 
\label{nsf_nccssg_dual}
\eeq
and hence 
\beq
\phi(\{G_{pl.,sd.}\}) = \sigma(\{G_{pl.,sd.}\}) \ .
\label{phi_sigma_sdrel}
\eeq
In particular, since the square lattice is planar and self-dual, we have 
\beq
\phi(sq) = \sigma(sq) \ , 
\label{phi_sigma_sq}
\eeq
so that the lower and upper bounds that we infer below for $\phi(sq)$ also 
hold for $\sigma(sq)$. For the triangular and honeycomb lattices, we obtain 
the relations
\beq
\phi(hc) = [\sigma(tri)]^{\nu(hc)} = [\sigma(tri)]^{1/2}  
\label{sigmatri_phihc}
\eeq
and
\beq
\sigma(hc) = [\phi(tri)]^{\nu(hc)} = [\phi(tri)]^{1/2} \ . 
\label{sigmahc_phitri}
\eeq

The Tutte polynomial is equivalent to the Whitney rank polynomial
\cite{whitney}, 
\beq
R(G,\xi,\eta) = \sum_{G' \subseteq G} \xi^{n(G)-k(G')} \eta^{c(G')}
\label{rwhitney}
\eeq
where $G'$ is a spanning subgraph of $G$ and $c(G')$ is the number of (linearly
independent) circuits on $G'$. Recall that $c(G')=e(G')+k(G')-n(G')$ and 
$n(G')=n(G)$. The equivalence is given by 
\beq
T(G,x,y) = (x-1)^{n(G)-k(G)} \, R(G,\xi,\eta) \ , 
\label{trrel}
\eeq
where
\beq
\xi = \frac{1}{x-1} \ , \quad \eta = y-1  \ . 
\label{xi_eta}
\eeq
(The variable $\xi$ in Eqs. (\ref{trrel}) and (\ref{xi_eta}) should not be 
confused with the symbol used for the
generic exponential growth constant in Eq. (\ref{ximin}).) 
The Tutte polynomial of a graph $G$ is also equivalent to a function of 
interest in statistical physics, namely the Potts model partition function, 
denoted $Z(G,q,v)$, which may be expressed as \cite{fk}:
\beq
Z(G,q,v) = \sum_{G' \subseteq G} q^{k(G')} v^{e(G')}  \ ,
\label{zcluster}
\eeq
where again, $G'$ is a spanning subgraph of $G$. The equivalence is given by 
\beq
Z(G,q,v) = (x-1)^{k(G)}(y-1)^{n(G)}T(G,x,y) \ , 
\label{zt}
\eeq
where 
\beq
x = 1 + \frac{q}{v} \ , \quad y = v+1 \ . 
\label{xy}
\eeq
so that $q=(x-1)(y-1)$.  Thus, one also has the equivalence 
\beq
Z(G,q,v) = q^{n(G)} R(G,\xi,\eta) \ , 
\label{zr}
\eeq
where 
\beq
\xi = \frac{v}{q} \ , \quad \eta = v \ . 
\label{xieta}
\eeq

\end{appendix}



\newpage


\begin{table}
  \caption{\footnotesize{Values of $\phi(\{G\})$ for
      the infinite-length limits of strip graphs of the square lattice with 
      width $L_y$ vertices and free (F) or periodic (P) transverse boundary
      conditions, $BC_y$. The infinite-length strip of a lattice $\Lambda$ 
      with width $L_y$ and given transverse boundary conditions is denoted 
      $\Lambda,(L_y)_{BC_y} \times \infty$; here, $\Lambda=sq$. As discussed
      in the text, $\phi(sq,(L_y)_{BC_y} \times \infty) =
      [\lambda_{sq,(L_y)_{BC_y}}(2,1)]^{1/L_y}$, and these values are 
      inferred to be lower bounds on $\phi(sq)$, with the values for 
      periodic $BC_y$ and the maximal $L_y$ being the most restrictive. 
      As defined in Eq. (\ref{ralpha}), $R_{\phi,sq,BC_y,\frac{L_y}{L_y-1}} 
      = \phi(sq,(L_y)_{BC_y} \times \infty)/
      \phi(sq,(L_y-1)_{BC_y} \times \infty)$. 
      Here and in subsequent tables, a blank entry means that the 
      evaluation is not applicable.}}
\begin{center}
\begin{tabular}{||l|l|l|l||}
\hline
BC$_y$ & $L_y$ & $\phi(sq,(L_y)_{BC_y} \times \infty)$ & 
$R_{\phi,sq,BC_y,\frac{L_y}{L_y-1} }$ \\ 
\hline
F & 1 & 2                      & \\ \hline
F & 2 & $1+\sqrt{3} = 2.73205081...$ & 1.36602540 \\ \hline
F & 3 & 3.02428923 & 1.10696669 \\ \hline
F & 4 & 3.18094706 & 1.05179988 \\ \hline
F & 5 & 3.27859286 & 1.03069709 \\ \hline
F & 6 & 3.34528558 & 1.02034187  \\ \hline
P & 2 & $\frac{\sqrt{15}+\sqrt{7}}{2} = 3.25936733..$ & \\ \hline
P & 3 & 3.53705348 & 1.08519634 \\ \hline
P & 4 & 3.62352967 & 1.02444865 \\ \hline
P & 5 & 3.65845648 & 1.00963889 \\ \hline
P & 6 & 3.67518338 & 1.00457212 \\ \hline
\end{tabular}
\end{center}
\label{lowerbounds_phi_sq_table}
\end{table}


\begin{table}
  \caption{\footnotesize{Upper bounds and their ratios for 
$\phi(sq)$ as functions of strip width $L_y$. The ratio
$R_{sq, \frac{L_y^2}{(L_y-1)(L_y+1)},F}(2,1)$ is defined in 
Eq. (\ref{ratio_upper}), where F denotes free transverse boundary conditions.}}
\begin{center}
\begin{tabular}{||l|l|l||}
\hline
$\frac{L_y+1}{L_y}$ & 
$\frac{\lambda_{sq,L_y+1,F}(2,1)}{\lambda_{sq,L_y,F}(2,1)}$ & 
$R_{sq, \frac{L_y^2}{(L_y-1)(L_y+1)},F} (2,1)$ \\ 
\hline
2/1 & $2+\sqrt{3}= 3.73205081..$ & \\ \hline
3/2 & 3.70588916 & 1.00705948 \\ \hline
4/3 & 3.70131286 & 1.00123640 \\ \hline
5/4 & 3.70008482 & 1.00033189 \\ \hline
6/5 & 3.69965942 & 1.00011498 \\ \hline
\end{tabular}
\end{center}
\label{upperbounds_phi_sq_table}
\end{table}


\begin{table}
  \caption{\footnotesize{Lower bounds and their ratios for 
$\phi(tri)$ as functions of strip width $L_y$.}} 
\begin{center}
\begin{tabular}{||l|l|l|l||}
\hline
BC$_y$ & $L_y$ & $\phi(tri,(L_y)_{BC_y} \times \infty)$ & 
$R_{\phi,tri,BC_y,\frac{L_y}{L_y-1},}$ \\ 
\hline
F & 2     & $\sqrt{2(3+2\sqrt{2})} = 3.41421356...$  & \\ \hline
F & 3     & 4.01637573 & 1.17636921 \\ \hline
F & 4     & 4.34758961 & 1.08246586 \\ \hline
F & 5     & 4.55702010 & 1.04817163 \\ \hline
F & 6     & 4.70139379 & 1.03168160 \\ \hline
P & 2     & $\frac{46+2\sqrt{505}}{2} = 4.76823893...$ & \\ \hline
P & 3     & 5.17697865 & 1.08572132 \\ \hline
P & 4     & 5.32006369 & 1.02763872 \\ \hline
P & 5     & 5.38485420 & 1.01217852 \\ \hline
\end{tabular}
\end{center}
\label{lowerbounds_phi_tri_table}
\end{table}


\begin{table}
  \caption{\footnotesize{Upper bounds and their ratios for 
$\phi(tri)$ as functions of strip width $L_y$.}} 
\begin{center}
\begin{tabular}{||l|l|l||}
\hline
$\frac{L_y+1}{L_y}$ & 
$\frac{\lambda_{tri, L_y+1,F} (2,1)}{\lambda_{tri, L_y,F}(2,1)}$ & 
$R_{tri, \frac{L_y^2}{(L_y-1)(L_y+1)},F}(2,1)$ \\ \hline
2/1 & $3+2\sqrt{2} = 5.82842712..$ & \\ \hline
3/2 & 5.55803958 & 1.04864801 \\ \hline
4/3 & 5.51430988 & 1.00793022 \\ \hline
5/4 & 5.50060617 & 1.00249131 \\ \hline
6/5 & 5.49484009 & 1.00104936 \\ \hline
\end{tabular}
\end{center}
\label{upperbounds_phi_tri_table}
\end{table}


\begin{table}
  \caption{\footnotesize{Lower bounds and their ratios for $\phi(hc)$ 
 as functions of strip width $L_y$.}}
\begin{center}
\begin{tabular}{||l|l|l|l||}
\hline
BC$_y$ & $L_y$ & $\phi(hc,(L_y)_{BC_y} \times \infty)$ & 
$R_{\phi,hc,\frac{L_y+1}{L_y} /\frac{L_y+2}{L_y},BC_y}$ \\ \hline
F   & 2  & $(16+4\sqrt{15})^{1/4} = 2.36891693..$ & \\ \hline
F   & 3  & 2.50613944 & 1.05792627 \\ \hline
F   & 4  & 2.57768156 & 1.02854674 \\ \hline
F   & 5  & 2.62158102 & 1.01703060 \\ \hline
F   & 6  & 2.65126155 & 1.01132162 \\ \hline
P   & 2  & $1+\sqrt{3} = 2.73205081$ & \\ \hline
P   & 4  & 2.79825703 & 1.02423316 \\ \hline
P   & 6  & 2.80378733 & 1.00197634 \\ \hline
\end{tabular}
\end{center}
\label{lowerbounds_phi_hc_table}
\end{table}
                                                                                

\begin{table}
  \caption{\footnotesize{Upper bounds and their ratios for 
$\phi(hc)$ as functions of strip width $L_y$.}}
\begin{center}
\begin{tabular}{||l|l|l||}
\hline
$(L_y+1)/L_y$ & $\Big [\lambda_{hc, L_y+1,F} (2,1) / \lambda_{hc, L_y,F}
(2,1) \Big ]^{1/2}$ & $R_{hc, \frac{L_y^2}{(L_y-1)(L_y+1)},F} (2,1)$ \\ \hline
2/1 & $\frac{\sqrt{6}+\sqrt{10}}{2} = 2.80588370..$ & \\ \hline
3/2 & 2.80489129 & 1.00035381 \\ \hline
4/3 & 2.80479655 & 1.00003378 \\ \hline
5/4 & 2.80478358 & 1.00000462 \\ \hline
6/5 & 2.80478142 & 1.00000077 \\ \hline
\end{tabular}
\end{center}
\label{upperbounds_phi_hc_table}
\end{table}


\begin{table}
  \caption{\footnotesize{Lower bounds on $\phi((4 \cdot 8^2))$ and their ratios,
      as functions of strip width $L_y$.}} 
\begin{center}
\begin{tabular}{||l|l|l|l||}
\hline
BC$_y$      & $L_y$ & $\phi((4 \cdot 8^2),(L_y)_{BC_y} \times \infty)$ & 
$R_{\phi,(4.8^2),\frac{L_y+1}{L_y} /\frac{L_y+2}{L_y},BC_y}$ \\ \hline
F   & 2  & $(478+2\sqrt{57057})^{1/8} = 2.35799035..$ & \\ \hline
F   & 3  & 2.49087484 & 1.056354974237... \\ \hline
F   & 4  & 2.56008993 & 1.027787462852... \\ \hline
F   & 5  & 2.60253811 & 1.016580737054... \\ \hline
F   & 6  & 2.63122712 & 1.011023473885... \\ \hline
P   & 2  & $1+\sqrt{3} = 2.73205081$ & \\ \hline
P   & 4  & 2.77638152 & 1.01622617 \\ \hline
P   & 6  & 2.77913516 & 1.00099181 \\ \hline
\end{tabular}
\end{center}
\label{lowerbounds_phi_488_table}
\end{table}


\begin{table}
  \caption{\footnotesize{Upper bounds on $\phi((4 \cdot 8^2))$ and their ratios,
      as functions of strip width $L_y$.}} 
\begin{center}
\begin{tabular}{||l|l|l||}
\hline
$(L_y+1)/L_y$ & $\Big [\lambda_{(4.8^2), L_y+1,F} (2,1) / 
                  \lambda_{(4.8^2), L_y,F} (2,1) \Big ]^{1/4}$ & 
$R_{(4.8^2),\frac{L_y^2}{(L_y-1)(L_y+1)},F} (2,1)$ \\ \hline
2/1 & $\frac{(487+2\sqrt{57057})^{1/4}}{2} = 2.78005925..$ & \\ \hline
3/2 & 2.77953194 & 1.00018971 \\ \hline
4/3 & 2.77949034 & 1.00001497 \\ \hline
5/4 & 2.77948671 & 1.00000131 \\ \hline
6/5 & 2.77948627 & 1.00000016 \\ \hline
\end{tabular}
\end{center}
\label{upperbounds_phi_488_table}
\end{table}


\begin{table}
  \caption{\footnotesize{Lower bounds on $\phi(kag)$ and their ratios,
      as functions of strip width $L_y$.}} 
\begin{center}
\begin{tabular}{||l|l|l|l||}
\hline
BC$_y$     & $L_y$ & $\phi(kag,(L_y)_{BC_y} \times \infty)$ &
$R_{\phi,kag,(L_y+1)/L_y, BC_y}$ \\ \hline
F    & 2   & $(97+\sqrt{8777})^{1/5} = 2.85800905..$ & \\ \hline 
F    & 3   & 3.12095363  & 1.09200271 \\ \hline
P    & 1   & $33^{1/3} = 3.20753433..$ & \\ \hline
P    & 2   & $\big(\frac{1991+19\sqrt{10545}}{2}\big)^{1/6} =
3.54091952..$ & 1.10393815 \\ \hline
P    & 3   & 3.59048515 & 1.01399796 \\ \hline
\end{tabular}
\end{center}
\label{lowerbounds_phi_kag_table}
\end{table}


\begin{table}
  \caption{\footnotesize{Upper bounds on $\phi(kag)$ and their ratios,
      as functions of strip width $L_y$.}} 
\begin{center}
\begin{tabular}{||l|l|l||}
\hline
$(L_y+1)/L_y$ & $\Big [\lambda_{kag, L_y+1,F} (2,1) / \lambda_{kag, L_y,F}
(2,1) \Big ]^{1/3}$ & 
$R_{kag,\frac{L_y^2}{(L_y-1)(L_y+1)},F} (2,1)$ \\ \hline
2/1 & $\big(\frac{97+\sqrt{8777}}{4}\big)^{1/3} = 3.62592933..$ & \\
3/2 & 3.6140446 & 1.00328848 \\ \hline
\end{tabular}
\end{center}
\label{upperbounds_phi_kag_table}
\end{table}


\begin{table}
  \caption{\footnotesize{Lower bounds on $\phi((3^3 \cdot 4^2))$ and their 
   ratios, as functions of strip width $L_y$. One can define different paths
      transverse to the longitudinal direction on a strip of this lattice
(see Fig. 1(a) in \cite{sti}). We list results for both choices.}} 
\begin{center}
\begin{tabular}{||l|l|l|l||}
\hline
BC$_y$ & $L_y$ & $\phi((3^3 \cdot 4^2), (L_y)_{BC_y} \times \infty)$ &
$R_{\phi,(3^3 \cdot 4^2),\frac{L_y+2}{L_y}/\frac{L_y+1}{L_y},BC_y}$ \\ 
\hline
F  & 3  & 3.49582205  & \\ \hline
F  & 5  & 3.88717879  & 1.11194985 \\ \hline
P  & 2  & $\sqrt{5}+\sqrt{3} = 3.96811879..$ & \\ \hline
P  & 4  & 4.42938725  & 1.11624361 \\ \hline
P  & 6  & 4.50622854  & 1.01734807 \\ \hline
\hline
F  & 2  & $(44+8\sqrt{30})^{1/4} = 3.06122777..$ & \\ \hline
F  & 3  & 3.49986242 & 1.143287167 \\ \hline
F  & 4  & 3.73916108 & 1.068373734 \\ \hline
F  & 5  & 3.88977485 & 1.040280095 \\ \hline
F  & 6  & 3.99328734 & 1.026611435 \\ \hline
P  & 2  & $\sqrt{123+\sqrt{15105}} = 3.95995902..$ & \\ \hline
P  & 3  & 4.30996446 & 1.088386127 \\ \hline
P  & 4  & 4.42859682 & 1.027525136 \\ \hline
P  & 5  & 4.48019516 & 1.011651170 \\ \hline
\end{tabular}
\end{center}
\label{lowerbounds_phi_33344_table}
\end{table}


\begin{table}
  \caption{\footnotesize{Upper bounds on $\phi((3^3 \cdot 4^2))$ 
and their ratios, as functions of strip width $L_y$. See caption to 
Table \ref{lowerbounds_phi_33344_table}.}} 
\begin{center}
\begin{tabular}{||l|l|l||}
\hline
$\frac{L_y+2}{L_y}$ or $\frac{L_y+1}{L_y}$ & $\Big [ \frac{\lambda_{(3^3.4^2),
      L_y+2/1,F} (2,1)}{\lambda_{(3^3.4^2), L_y,F} (2,1)} \Big ]^{1/2}$ &
$R_{(3^3.4^2),\frac{L_y^2}{(L_y-2)(L_y+2)}/\frac{L_y^2}{(L_y-1)(L_y+1)},F}
 (2,1)$ \\ \hline
3/1 & 4.62177690 & \\ \hline
5/3 & 4.55787399 & 1.01402033 \\ \hline
\hline
2/1 & $\sqrt{5}+\sqrt{6} = 4.68555772...$ & \\ \hline
3/2 & 4.57468959 & 1.02423512 \\ \hline
4/3 & 4.55977887 & 1.00327005 \\ \hline
5/4 & 4.55539056 & 1.00096332 \\ \hline
6/5 & 4.55366469 & 1.00037901 \\ \hline
\end{tabular}
\end{center}
\label{upperbounds_phi_33344_table}
\end{table}


\begin{table}
  \caption{\footnotesize{Lower bounds and their ratios for 
$\phi((3^2 \cdot 4 \cdot 3 \cdot 4))$ as 
functions of strip width $L_y$.}} 
\begin{center}
\begin{tabular}{||l|l|l|l||}
\hline
BC$_y$  & $L_y$ & $\phi((3^2 \cdot 4 \cdot 3 \cdot 4), (L_y)_{BC_y} 
\times \infty)$ &
$R_{\phi,(3^2 \cdot 4 \cdot 3 \cdot 4), 
\frac{L_y+1}{L_y}/\frac{L_y+2}{L_y},BC_y}$ \\ 
\hline
F  & 2  & $(44+8\sqrt{30})^{1/4} = 3.06122777..$ & \\ \hline
F  & 3  & 3.50500542 & 1.14496721  \\ \hline
F  & 4  & 3.74646778 & 1.06889072  \\ \hline
F  & 5  & 3.89838787 & 1.04055022  \\ \hline
F  & 6  & 4.00278463 & 1.02677947  \\ \hline
P  & 2  & $\sqrt{5}+\sqrt{3} = 3.968118785..$ & \\ \hline
P  & 4  & 4.43763851 & 1.11832300 \\ \hline
\end{tabular}
\end{center}
\label{lowerbounds_phi_33434_table}
\end{table}


\begin{table}
  \caption{\footnotesize{Upper bounds and their ratios for 
$\phi((3^2 \cdot 4 \cdot 3 \cdot 4))$ as functions of strip width $L_y$.}} 
\begin{center}
\begin{tabular}{||l|l|l||}
\hline
$(L_y+1)/L_y$&$\sqrt{ \frac{\lambda_{(3^2 \cdot 4 \cdot 3 \cdot 4),L_y+1,F}
    (2,1)}{\lambda_{(3^2 \cdot 4 \cdot 3 \cdot 4), L_y,F} (2,1)} }$ & 
$R_{(3^2 \cdot 4 \cdot 3 \cdot 4), 
\frac{L_y^2}{(L_y-1)(L_y+1)},F} (2,1)$ \\ \hline
2/1 & $\sqrt{5}+\sqrt{6} = 4.68555772..$ & \\ \hline
3/2 & 4.59488654 & 1.01973306 \\ \hline
4/3 & 4.57532478 & 1.00427549 \\ \hline
5/4 & 4.57022128 & 1.00111668 \\ \hline
6/5 & 4.56823149 & 1.00043557 \\ \hline
\end{tabular}
\end{center}
\label{upperbounds_phi_33434_table}
\end{table}



\begin{table}
  \caption{\footnotesize{Lower bounds and their ratios for $\sigma(sq)$ as 
functions of strip width $L_y$.}} 
\begin{center}
\begin{tabular}{||l|l|l|l||}
\hline
BC$_y$ & $L_y$ & $\sigma(sq,(L_y)_{BC_y} \times \infty)$ & 
$R_{\sigma,sq, \frac{L_y}{L_y-1},BC_y}$ \\ 
\hline
F  & 1  & 1                 & \\ \hline
F  & 2  & $\frac{\sqrt{10+2\sqrt{17}}}{2} = 2.13577921..$ & 2.13577921..\\ 
\hline
F  & 3  & 2.62742787    & 1.23019639 \\ \hline
F  & 4  & 2.88792764    & 1.09914631 \\ \hline
F  & 5  & 3.04750858    & 1.05525794 \\ \hline
F  & 6  & 3.15487018    & 1.03522930 \\ \hline
P  & 2  & $\frac{\sqrt{15}+\sqrt{7}}{2} = 3.25936733..$ & \\ \hline
P  & 3  & 3.53705348    & 1.08519634 \\ \hline
P  & 4  & 3.62352967    & 1.02444865 \\ \hline
P  & 5  & 3.65845648    & 1.00963889 \\ \hline
P  & 6  & 3.67518338    & 1.00457212 \\ \hline
\end{tabular}
\end{center}
\label{lowerbounds_sigma_sq_table}
\end{table}


\begin{table}
  \caption{\footnotesize{Upper bounds and their ratios for $\sigma(sq)$ as 
functions of strip width $L_y$.}} 
\begin{center}
\begin{tabular}{||l|l|l||}
\hline
$(L_y+1)/L_y$ & $\lambda_{sq,L_y+1,F}(1,2)/\lambda_{sq,L_y,F}(1,2)$ & 
$R_{sq, \frac{L_y^2}{(L_y-1)(L_y+1)},F}(1,2)$ \\ \hline
2/1 & $\frac{5+\sqrt{17}}{2} = 4.56155281..$ & \\ \hline
3/2 & 3.97630508 & 1.14718381 \\ \hline
4/3 & 3.83488921 & 1.03687613 \\ \hline
5/4 & 3.77902232 & 1.01478342 \\ \hline
6/5 & 3.75114866 & 1.00743070 \\ \hline
\end{tabular}
\end{center}
\label{upperbounds_sigma_sq_table}
\end{table}


\begin{table}
  \caption{\footnotesize{Lower bounds and their ratios for $\sigma(tri)$ as 
functions of strip width $L_y$.}}
\begin{center}
\begin{tabular}{||l|l|l|l||}
\hline
BC$_y$   & $L_y$ & $\sigma(tri,(L_y)_{BC_y} \times \infty)$ &
$R_{\sigma,tri,(L_y+1)/L_y,BC_y}$ \\ 
\hline
F  & 2  & $\sqrt{6+4\sqrt{2}} = 3.41421356..$ & \\ \hline
F  & 3  & 4.65472093 & 1.36333620 \\ \hline
F  & 4  & 5.35640463 & 1.15074668 \\ \hline
P  & 5  & 5.80398594 & 1.08356003 \\ \hline
P  & 6  & 6.11427423 & 1.05346124 \\ \hline
P  & 2  & $\sqrt{29+\sqrt{817}} = 7.58836029..$ & \\ \hline
P  & 3  & 7.80037170 & 1.02793903 \\ \hline
P  & 4  & 7.84674402 & 1.00594489 \\ \hline
P  & 5  & 7.85992934 & 1.00168036 \\ \hline
\end{tabular}
\end{center}
\label{lowerbounds_sigma_tri_table}
\end{table}

\clearpage 


\begin{table}
  \caption{\footnotesize{Upper bounds, their ratios relative to the exact
      $\sigma(tri)$, and ratios of adjacent bounds, as 
functions of strip width $L_y$.}} 
\begin{center}
\begin{tabular}{||l|l|l||}
\hline
$(L_y+1)/L_y$ & $\lambda_{tri,L_y+1,F}(1,2)/\lambda_{tri,L_y,F}(1,2)$ & 
$R_{tri, \frac{L_y^2}{(L_y-1)(L_y+1)},F}(1,2)$ \\ 
\hline
2/1 & $6+4\sqrt{2} = 11.65685425..$ & \\ \hline
3/2 & 8.65166268 & 1.34735422 \\ \hline
4/3 & 8.16230020 & 1.05995399 \\ \hline
5/4 & 8.00088909 & 1.02017415 \\ \hline
6/5 & 7.93300485 & 1.00855719 \\ \hline
\end{tabular}
\end{center}
\label{upperbounds_sigma_tri_table}
\end{table}


\begin{table}
  \caption{\footnotesize{Lower bounds and their ratios for $\sigma(hc)$ as 
functions of strip width $L_y$.}}
\begin{center}
\begin{tabular}{||l|l|l|l||}
\hline
BC$_y$    & $L_y$ & $\sigma(hc,(L_y)_{BC_y} \times \infty)$ &
$R_{\sigma,hc,\frac{L_y+1}{L_y} /\frac{L_y+2}{L_y},BC_y}$ \\ 
\hline
F   & 2  & $\big ( \frac{7+\sqrt{41}}{2} \big )^{1/4} = 1.60895542..$ & \\ 
\hline
F   & 3  & 1.84524123 & 1.14685665 \\ \hline
F   & 4  & 1.96759470 & 1.06630758 \\ \hline
F   & 5  & 2.04197649 & 1.03780341 \\ \hline
F   & 6  & 2.09186520 & 1.02443158 \\ \hline
P   & 2  & $\frac{10+2\sqrt{17}}{2} = 2.13577921..$ & \\ \hline
P   & 4  & 2.29347361 & 1.07383460 \\ \hline
P   & 6  & 2.32235509 & 1.01259290 \\ \hline
\end{tabular}
\end{center}
\label{lowerbounds_sigma_hc_table}
\end{table}


\begin{table}
  \caption{\footnotesize{Upper bounds and their ratios for $\sigma(hc)$ as 
functions of strip width $L_y$.}}
\begin{center}
\begin{tabular}{||l|l|l||}
\hline
$(L_y+1)/L_y$ & $\Big [ \lambda_{hc, L_y+1,F} (1,2) /\lambda_{hc, L_y,F}
(1,2) \Big ]^{1/2}$ & 
$R_{hc, \frac{L_y^2}{(L_y-1)(L_y+1)},F} (1,2)$ \\ \hline
2/1 & $\frac{\sqrt{14+2\sqrt{41}}}{2} = 2.588737553078...$ & \\ \hline
3/2 & 2.42700921 & 1.06663689 \\ \hline
4/3 & 2.38552036 & 1.01739195 \\ \hline
5/4 & 2.36870574 & 1.00709866 \\ \hline
6/5 & 2.36019825 & 1.00360457 \\ \hline
\end{tabular}
\end{center}
\label{upperbounds_sigma_hc_table}
\end{table}


\begin{table}
  \caption{\footnotesize{Lower bounds and their ratios for 
$\sigma((4 \cdot 8^2))$ as functions of strip width $L_y$.}}
\begin{center}
\begin{tabular}{||l|l|l|l||}
\hline
BC$_y$   & $L_y$ & $\sigma((4 \cdot 8^2), (L_y)_{BC_y} \times \infty)$ &
$R_{\sigma,(4 \cdot 8^2), \frac{L_y+1}{L_y} /\frac{L_y+2}{L_y}, BC_y}$ \\ 
\hline
F  & 2  & $\big ( \frac{41+3\sqrt{185}}{2} \big)^{1/8} = 1.59026075..$ & \\ 
\hline
F  & 3  & 1.82207863  & 1.14577350 \\ \hline
F  & 4  & 1.94327804  & 1.06651711 \\ \hline
F  & 5  & 2.01743612  & 1.03816133 \\ \hline
F  & 6  & 2.06740176  & 1.02476690 \\ \hline
P  & 2  & $\frac{10+2\sqrt{17}}{2} = 2.13577921..$ & \\ \hline
P  & 4  & 2.27644959  & 1.06586373 \\ \hline
P  & 6  & 2.30261139  & 1.01149237 \\ \hline
\end{tabular}
\end{center}
\label{lowerbounds_sigma_488_table}
\end{table}


\begin{table}
  \caption{\footnotesize{Upper bounds and their ratios for 
$\sigma((4 \cdot 8^2))$ as functions of strip width $L_y$.}}
\begin{center}
\begin{tabular}{||l|l|l||}
\hline
$(L_y+1)/L_y$ & $\Big [ \lambda_{(4.8^2), L_y+1,F} (1,2) /
\lambda_{(4.8^2), L_y,F} (1,2) \Big ]^{1/4}$ & 
$R_{(4.8^2), \frac{L_y^2}{(L_y-1)(L_y+1)},F} (1,2)$ \\ 
\hline
2/1 & $\big ( \frac{(41+3\sqrt{185}}{2} \big)^{1/4} = 2.52892927..$ & \\ 
\hline
3/2 & 2.39201919 & 1.05723620 \\ \hline
4/3 & 2.35742796 & 1.01467329 \\ \hline
5/4 & 2.34346889 & 1.00595658 \\ \hline
6/5 & 2.33641686 & 1.00301831 \\ \hline
\end{tabular}
\end{center}
\label{upperbounds_sigma_488_table}
\end{table}


\begin{table}
\caption{\footnotesize{Lower bounds and their ratios for 
$\sigma(kag)$ as functions of strip width $L_y$.}}
\begin{center}
\begin{tabular}{||l|l|l|l||}
\hline
BC$_y$  & $L_y$ & $\sigma(kag,(L_y)_{BC_y} \times \infty)$ & 
$R_{\sigma,kag, (L_y+1)/L_y,BC_y}$ \\ 
\hline
F   & 2  & $(40+12\sqrt{10})^{1/5} = 2.38979281..$ & \\
\hline
F   & 3  & 2.85572120 & 1.19496602 \\ \hline
P   & 1  & $32^{1/3} = 3.17480210..$ & \\ \hline
P   & 2  & $(1056+128\sqrt{66})^{1/6} = 3.57734613..$ & 1.12679342 \\ \hline
P   & 3  & 3.64470247  & 1.01882858 \\ \hline
\end{tabular}
\end{center}
\label{lowerbounds_sigma_kag_table}
\end{table}


\begin{table}
\caption{\footnotesize{Upper bounds and their ratios for 
$\sigma(kag)$ as functions of strip width $L_y$.}}
\begin{center}
\begin{tabular}{||l|l|l||}
\hline
$(L_y+1)/L_y$ & $\Big [ \lambda_{kag, L_y+1,F} (1,2) /
\lambda_{kag, L_y,F} (1,2) \Big ]^{1/3}$ & 
$R_{kag,\frac{L_y^2}{(L_y-1)(L_y+1)},F} (1,2)$ \\ 
\hline
2/1 & $(40+12\sqrt{10})^{1/3} = 4.27169679..$ & \\ \hline
3/2 & 3.84274644 & 1.11162598 \\ \hline
\end{tabular}
\end{center}
\label{upperbounds_sigma_kag_table}
\end{table}


\begin{table}
\caption{\footnotesize{Lower bounds and their ratios for 
$\sigma((3^3 \cdot 4^2))$ as functions of strip width $L_y$.}}
\begin{center}
\begin{tabular}{||l|l|l|l||}
\hline
BC$_y$ & $L_y$ & $\sigma((3^3 \cdot 4^2),(L_y)_{BC_y} \times \infty)$ &
$R_{\sigma,(3^3 \cdot 4^2), \frac{L_y+2}{L_y}/\frac{L_y+1}{L_y},BC_y}$ \\ 
\hline
F  & 3 & 3.51850132 & \\ \hline
F  & 5 & 4.24280788 & 1.20585656 \\ \hline
P  & 2 & $\sqrt{13+\sqrt{161}} = 5.06839003..$ & \\ \hline
P  & 4 & 5.40602876 & 1.06661656 \\ \hline
P  & 6 & 5.44463590 & 1.00714150 \\ \hline \hline
F  & 2 & $(27+\sqrt{721})^{1/4} = 2.70893969..$ & \\ \hline
F  & 3 & 3.51703426  & 1.29830659 \\ \hline
F  & 4 & 3.96327800  & 1.12688069 \\ \hline
F  & 5 & 4.24306553  & 1.07059498 \\ \hline
F  & 6 & 4.43422383  & 1.04505193 \\ \hline
P  & 2 & $\sqrt{313+\sqrt{97873}} = 5.00169236...$ & \\ \hline
P  & 3 & 5.30268205  & 1.06017757 \\ \hline
P  & 4 & 5.39237466  & 1.01691457 \\ \hline
P  & 5 & 5.42627704  & 1.00628710 \\ \hline
\end{tabular}
\end{center}
\label{lowerbounds_sigma_33344_table}
\end{table}


\begin{table}
\caption{\footnotesize{Upper bounds and their ratios for 
$\sigma((3^3 \cdot 4^2))$ as functions of strip width $L_y$.}}
\begin{center}
\begin{tabular}{||l|l|l||}
 \hline
$\frac{L_y+2}{L_y}$ or $\frac{L_y+1}{L_y}$ & $\Big [
\frac{\lambda_{ (3^3 \cdot 4^2),
L_y+2/1,F} (1,2)} {\lambda_{ (3^3 \cdot 4^2), L_y,F} (1,2)} \Big ]^{1/2}$ &
$R_{(3^3 \cdot 4^2),\frac{L_y^2}{(L_y-2)(L_y+2)}/
\frac{L_y^2}{(L_y-1)(L_y+1)},F} (1,2)$ \\ 
\hline
3/1 & 6.59988817 & \\ \hline
5/3 & 5.61819539 & 1.17473454 \\ \hline\hline
2/1 & $\frac{\sqrt{27+\sqrt{721}}}{2} = 7.33835425$ & \\ \hline
3/2 & 5.92831296 & 1.23784866 \\ \hline
4/3 & 5.67137481 & 1.04530439 \\ \hline
5/4 & 5.57417435 & 1.01743764 \\ \hline
6/5 & 5.52722284 & 1.00849459 \\ \hline
\end{tabular}
\end{center}
\label{upperbounds_sigma_33344_table}
\end{table}


\begin{table}
\caption{\footnotesize{Lower bounds and their ratios for 
$\sigma((3^2 \cdot 4 \cdot 3 \cdot 4))$ as functions of strip width $L_y$.}}
\begin{center}
\begin{tabular}{||l|l|l|l||}
\hline
BC$_y$ & $L_y$ & 
$\sigma((3^2 \cdot 4 \cdot 3 \cdot4), (L_y)_{BC_y} \times \infty)$ &
$R_{\sigma,(3^2 \cdot 4 \cdot 3 \cdot 4), 
\frac{L_y+1}{L_y}/\frac{L_y+2}{L_y},BC_y}$ \\ 
\hline
F   & 2  & $(27+\sqrt{721})^{1/4} = 2.70893969..$ & \\ \hline
F   & 3  & 3.52704267 & 1.30200118 \\ \hline
F   & 4  & 3.97204751 & 1.12616940 \\ \hline
F   & 5  & 4.25003524 & 1.06998600 \\ \hline
F   & 6  & 4.43971476 & 1.04463010 \\ \hline
P   & 2  & $\sqrt{13+\sqrt{161}} = 5.06839003$ & \\ \hline
P   & 4  & 5.40726946 & 1.06686136 \\ \hline
\end{tabular}
\end{center}
\label{lowerbounds_sigma_33434_table}
\end{table}


\begin{table}
\caption{\footnotesize{Upper bounds and their ratios for 
$\sigma((3^2 \cdot 4 \cdot 3 \cdot 4))$ as functions of strip width $L_y$.}}
\begin{center}
\begin{tabular}{||l|l|l||}
\hline 
$\frac{L_y+1}{L_y}$ & $\Big [ 
\lambda_{ (3^2 \cdot 4 \cdot 3 \cdot 4),L_y+1,F}(1,2)/
  \lambda_{ (3^2 \cdot 4 \cdot 3 \cdot 4), L_y,F} (1,2) \Big ]^{1/2}$ & 
$R_{(3^2 \cdot 4 \cdot 3 \cdot 4), \frac{L_y^2}{(L_y-1)(L_y+1)},F} (1,2)$ \\ 
\hline
2/1 & $\sqrt{27+\sqrt{721}} = 7.33835425..$ & \\ \hline
3/2 & 5.97906767 & 1.22734089 \\ \hline
4/3 & 5.67316738 & 1.05392055 \\ \hline
5/4 & 5.57063774 & 1.01840537 \\ \hline
6/5 & 5.52290732 & 1.00864226 \\ \hline
\end{tabular}
\end{center}
\label{upperbounds_sigma_33434_table}
\end{table}


\begin{table}
\caption{\footnotesize{Lower and upper bounds on 
    $\phi(\Lambda)$, denoted $\phi_\ell(\Lambda)$ and $\phi_u(\Lambda)$, 
    and on $\sigma(\Lambda)$, denoted 
    $\sigma_\ell(\Lambda)$ and $\sigma_u(\Lambda)$, for the lattices
    $\Lambda$ analyzed here. The lattices are listed
    in order of increasing vertex degree $\Delta(\Lambda)$. 
    See text for further discussion.}}
\begin{center}
\begin{tabular}{|c|c|c|c|c|c|} \hline\hline
$\Lambda$ & $\Delta(\Lambda)$ & $\phi_\ell(\Lambda)$ & $\phi_u(\Lambda)$ &
$\sigma_\ell(\Lambda)$ & $\sigma_u(\Lambda)$ \\
\hline
$(4 \cdot 8^2)$  & 3 & 2.779135 & 2.779486 & 2.302611 & 2.336417  \\ \hline 
$(6^3)=$ hc      & 3 & 2.803787 & 2.804781 & 2.322355 & 2.344107  \\ \hline 
$(4^4)=$ sq      & 4 & 3.675183 & 3.699659 & 3.675183 & 3.699659  \\ \hline 
$(3\cdot 6\cdot 3\cdot 6)$
                 & 4 & 3.590485 & 3.614045 & 3.644702 & 3.842746  \\ \hline 
$(3^3 \cdot 4^2)$& 5 & 4.506228 & 4.553665 & 5.444636 & 5.527223  \\ \hline 
$(3^2 \cdot 4 \cdot 3 \cdot 4)$
                 & 5 & 4.437638 & 4.568231 & 5.407269 & 5.522907  \\ \hline
$(3^6)=$ tri     & 6 & 5.393333 & 5.494840 & 7.861223 & 7.866799  \\ 
\hline\hline
\end{tabular}
\end{center}
\label{phi_sigma_bounds_table}
\end{table}

\begin{table}
\caption{\footnotesize{Approximate values $\phi_{app}(\Lambda)$ and 
   $\sigma_{app}(\Lambda)$, as defined in Eq. (\ref{xi_approx}), 
   for the lattices $\Lambda$ analyzed here.}}
\begin{center}
\begin{tabular}{|c|c|c|c|} \hline\hline
$\Lambda$ & $\Delta(\Lambda)$ & $\phi_{app}(\Lambda)$ & 
$\sigma_{app}(\Lambda)$ \\
\hline
$(4 \cdot 8^2)$ & 3 & $2.77931 \pm 0.00018$ & $2.3195 \pm 0.017$ \\ \hline 
$(6^3)=$ hc     & 3 & $2.80428 \pm 0.00050$ & $2.333 \pm 0.011$  \\ \hline 
$(4^4)=$ sq     & 4 & $3.687 \pm 0.012$     & $3.687 \pm 0.012$  \\ \hline 
$(3\cdot 6\cdot 3\cdot 6)$
                & 4 & $3.602 \pm 0.012$   & $3.74 \pm 0.10$   \\ \hline 
$(3^3 \cdot 4^2)$& 5 & $4.530 \pm 0.024$  & $5.486 \pm 0.041$ \\ \hline 
$(3^2 \cdot 4 \cdot 3 \cdot 4)$
                 & 5 & $4.503 \pm 0.065$   & $5.465 \pm 0.058$  \\ \hline
$(3^6)=$ tri     & 6 & $5.444 \pm 0.051$     & $7.864 \pm 0.0028$  \\ 
\hline\hline
\end{tabular}
\end{center}
\label{phi_sigma_approx_table}
\end{table}


\end{document}